\documentclass[preprint,12pt,review]{elsarticle}

\usepackage{amssymb}
\usepackage{amsthm}
\usepackage{amsmath}
\usepackage{mathrsfs}
\usepackage{graphicx}
\usepackage{epstopdf}
\usepackage{float}
\usepackage{caption}
\usepackage{subcaption}
\usepackage{bm}
\usepackage{extarrows} 
\usepackage{bbm}
\usepackage{soul}
\usepackage{booktabs}
\usepackage{makecell}
\usepackage[margin=2cm]{geometry}
\usepackage{longtable}
\usepackage{xcolor} 
\usepackage[colorlinks, linkcolor=blue, urlcolor=blue, citecolor=blue]{hyperref}
\usepackage{cleveref}
\usepackage{threeparttable}
\usepackage{adjustbox}
\biboptions{sort&compress}

\usepackage{lineno}

\journal{Engineering Fracture Mechanics}

\makeatletter
\def\@author#1{\g@addto@macro\elsauthors{\normalsize%
    \def\baselinestretch{1}%
    \upshape\authorsep#1\unskip\textsuperscript{%
      \ifx\@fnmark\@empty\else\unskip\sep\@fnmark\let\sep=,\fi
      \ifx\@corref\@empty\else\unskip\sep\@corref\let\sep=,\fi
      }%
    \def\authorsep{\unskip,\space}%
    \global\let\@fnmark\@empty
    \global\let\@corref\@empty  
    \global\let\sep\@empty}%
    \@eadauthor={#1}
}
\makeatother

\begin{document}

\begin{frontmatter}



\title{On the fracture mechanics validity of small scale tests}


\author{Chuanjie Cui\fnref{Ox}}

\author{Livia Cupertino-Malheiros\fnref{ICC}}

\author{Ziyao Xiong\fnref{ICP}}

\author{Emilio Mart\'{\i}nez-Pa\~neda\corref{cor1}\fnref{Ox}}
\ead{emilio.martinez-paneda@eng.ox.ac.uk}

\address[Ox]{Department of Engineering Science, University of Oxford, Oxford OX1 3PJ, UK}

\address[ICC]{Department of Civil and Environmental Engineering, Imperial College London, London SW7 2AZ, UK}

\address[ICP]{Department of Physics, Imperial College London, London SW7 2AZ, UK}

\cortext[cor1]{Corresponding author.}


\begin{abstract}
There is growing interest in conducting small-scale tests to gain additional insight into the fracture behaviour of components across a wide range of materials. For example, micro-scale mechanical tests inside of a microscope (\emph{in situ}) enable direct, high-resolution observation of the interplay between crack growth and microstructural phenomena (e.g., dislocation behaviour or the fracture resistance of a particular interface), and sub-size samples are increasingly used when only a limited amount of material is available. However, to obtain quantitative insight and extract relevant fracture parameters, the sample must be sufficiently large for a $J$- (HRR) or a $K$-field to exist. We conduct numerical and semi-analytical studies to map the conditions (sample geometry, material) that result in a valid, quantitative fracture experiment. Specifically, for a wide range of material properties, crack lengths and sample dimensions, we establish the maximum value of the $J$-integral where an HRR field ceases to exist (i.e., the maximum $J$ value at which fracture must occur for the test to be valid, $J_\mathrm{max}$). Maps are generated to establish the maximum valid $J$ value ($J_\mathrm{max}$) as a function of yield strength, strain hardening and minimum sample size. These maps are then used to discuss the existing experimental literature and provide guidance on how to conduct quantitative experiments. Finally, our study is particularised to the analysis of metals that have been embrittled due to hydrogen exposure. The response of relevant materials under hydrogen-containing environments are superimposed on the aforementioned maps, determining the conditions that will enable quantitative insight. \\

\end{abstract}


\end{frontmatter}


\section{Introduction}
\label{Sec:Intro} 

\begin{figure}[H]
\centering
\noindent\makebox[\textwidth]{%
\includegraphics[scale=0.5]{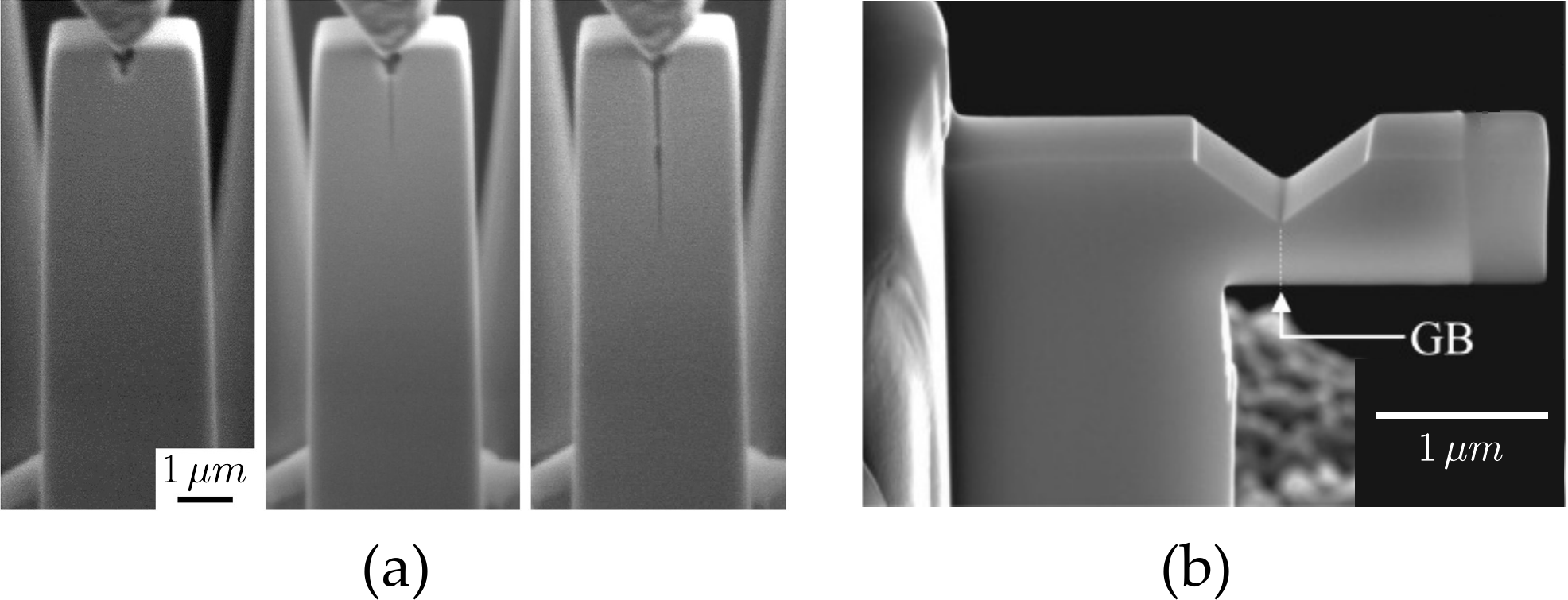}}
\caption{SEM images of two typical small scale testing specimens: (a) double cantilever beam (DCB), and (b) notched cantilever beam specimen. Adapted from \cite{wang_shear_2024,takahashi_direct_2016}.}
\label{fig:Exp_photo}
\end{figure}

Recent years have seen a surge in the development of small scale fracture experiments \cite{martinez2016fracture,jayaram_small-scale_2022,patel2023toughness,patel2024rethinking}. These are needed to gain insight into the structural integrity of micron-size components such as micro-electromechanical systems (MEMS) \cite{modlinski_alcumgmn_2008}, microelectronic components \cite{xiao2008interfacial} and thin film applications \cite{wu_micro-scale_2013}. Micro-scale fracture tests have also been conducted to obtain high-resolution observations of fracture processes inside electron microscopes (SEM and TEM). Such \emph{in situ} experiments have, for example, been used to gain insight into the fracture properties of ceramics at the length scale of individual interfaces \cite{sernicola_situ_2017,gavalda-diaz_observing_2024}, setting the basis for the development of fracture-resistant microstructures. Similarly, \emph{in situ} micromechanical fracture experiments have been conducted to enable real-time observation of crack initiation and growth in metals \cite{connolley_review_2005,zhang_situ_2024,reiser_metals_2020,Shahini2024,Kaveh2025} and its interaction with microstructural features, such as grain boundaries \cite{malyar_synthesis_2019,wang_measurement_2022} and phase interfaces \cite{chan_micro-scale_2016}. The approach has also been extended to include environmental effects, encompassing thermomechanical \cite{kang_situ_2017,morris_role_2018,fritz_development_2017}, electromechanical \cite{lee_electromechanical_2010,wang_mechanical_2025}, and chemomechanical \cite{takahashi_direct_2016,deng_-situ_2017,asadipoor_macro-_2020,tao_review_2023} phenomena. Multiple test configurations have been employed with the two shown in Fig. \ref{fig:Exp_photo} being arguably the most popular: the double cantilever beam splitting test (Fig. \ref{fig:Exp_photo}a) and the notched micro-cantilever beam specimen (Fig. \ref{fig:Exp_photo}b). At a higher scale (mm), sub-sized tests have been proposed to characterise fracture behaviour when the available material is limited as in (e.g.) a reactor pressure vessel or when sampling small regions of material (like the heat affected zone of a weld) \cite{cicero2020fracture,madi2024mechanical}. While the utility and additional insight of small scale tests are clear, their ability to provide quantitative fracture properties can be compromised by the small sample dimensions (relative to the plastic zone size or the fracture process zone size).

For a fracture test to be valid, the \emph{principle of similitude} must be met. That is, the fracture toughness measured must be independent of the sample geometry and thus applicable to both lab specimens and engineering components. For this condition to be satisfied, the stress distribution ahead of the crack tip must be described by either the linear elastic $r^{-1/2}$ singularity, which would render a valid $K_{Ic}$ estimate, or the Hutchinson-Rice-Rosengren (HRR) \cite{hutchinson_singular_1968,rice_plane_1968} singularity $r^{N/(N+1)}$, for a valid $J_{Ic}$ estimate. Here, $r$ is the distance to the crack tip, for a polar coordinate system centred at the crack tip, and $N$ is the strain hardening exponent ($0<N<1$). If a $K$ or a $J$-field exist ahead of the crack in the tested sample, the stress state is uniquely characterised and the values of $K_{Ic}$ or $J_{Ic}$ estimated are universal material properties. At sufficiently low applied loads, both a $K$-field and a $J$-field exist (small scale yielding conditions), but as the plastic or fracture process zone develops and becomes sufficiently large (relative to the sample size), the elastic singularity no longer applies. For a $K$-field to exist at the moment of fracture, the condition,
\begin{equation}\label{eq:LEFMvalidity}
    a \, , \,\, \left( W - a \right) > 2.5 \left( \frac{K_{Ic}}{\sigma_Y} \right)^2
\end{equation}
\noindent must be fulfilled. Where $K_{Ic}$ is the fracture toughness, $\sigma_Y$ is the yield strength, $a$ is the length of the initial crack, and $W$ is the width of the sample, such that ($W-a$) is the length of the ligament ahead of the crack. A $J$-field continues to exist in the presence of large scale crack tip plasticity but eventually disappears if the load and sample size conditions are such that large strain phenomena fully govern crack tip behaviour. For a $J$-based analysis to be valid, the following condition must be satisfied:
\begin{equation}\label{eq:HRRvalidity}
     a \, , \,\, \left( W - a \right) > M \frac{J_{Ic}}{\sigma_Y}
\end{equation}
\noindent where $J_{Ic}$ is the critical value of $J$ at fracture (the material toughness) and $M$ is a coefficient that is typically taken to be equal to 25 (see Section \ref{Sec:HRR} for details). Considering the plane stress relationship between $J_{Ic}$ and $K_{Ic}$ ($J_{Ic}=K_{Ic}^2/E$, where $E$ is the Young's modulus), it can be readily seen that the right side of Eq. (\ref{eq:HRRvalidity}) is $10 \sigma_Y /E$ times smaller than that of Eq. (\ref{eq:LEFMvalidity}), with $\sigma_Y/E$ being typically within the range $10^{-3}-10^{-2}$. In other words, the conditions required for the existence of a $J$-field are much less demanding, enabling the use of samples with ligament sizes that are 10 to 100 times shorter. Hence, the focus of this work will be on valid $J$-tests. It is of interest to note that Eqs. (\ref{eq:LEFMvalidity})-(\ref{eq:HRRvalidity}) are simple, practical semi-analytical estimates that do not take into consideration the influence of relevant material phenomena, such as strain hardening. We shall conduct finite element studies to rigorously characterise the conditions in which a small scale fracture test can be quantitative; i.e., the conditions in which a $J$-field exists.\\

In this work, we combine semi-analytical and numerical analysis to establish the regimes of validity of small scale fracture tests. Detailed large strain, elastic-plastic finite element calculations are conducted to determine - for a wide range of materials and sample dimensions - the value of $J_\mathrm{max}$, the maximum value of the $J$-integral where an HRR field ceases to exist (i.e., the maximum $J$ value at which fracture must occur to obtain a valid test). This enables constructing `validity maps' that provide the sample dimensions that render a valid fracture test, as a function of the material properties ($J_{Ic}$, $\sigma_Y$, $N$). These maps are novel, to the best of the authors' knowledge, and are of notable importance as they not only allow experimentalists to readily determine the appropriate sample dimensions but also allow us to evaluate the validity of existing literature data. Finally, we particularise our study to the emerging area of hydrogen embrittlement \cite{djukic_synergistic_2019,chen_hydrogen_2024}, conducting an extensive literature review to superimpose on the aforementioned $J_\mathrm{max}$ maps the fracture toughness in hydrogen-containing environments, for relevant materials and hydrogen environments. Once again, this enables us to discuss existing findings and bring new insight into the possibilities of conducting direct, quantitative experiments that could bring novel understanding into this challenging phenomenon.\\

The remainder of this manuscript is structured as follows. First, in Section \ref{Sec:Analytical}, we briefly introduce the theoretical foundation of the HRR singularity field and build on this to establish simple, semi-analytical estimates of valid sample dimensions as a function of the material fracture toughness. This allows us to provide the first validity map, which sets the basis for the subsequent discussion. This map is then refined through detailed finite element calculations which, among others, take into consideration the influence of strain hardening and testing configuration. The modelling framework is presented in Section \ref{Sec:Model}, together with the details of the two geometries considered: the double cantilever beam (Fig. \ref{fig:Exp_photo}a) and the notched micro-cantilever beam (Fig. \ref{fig:Exp_photo}b). The numerical results are presented and discussed in Section \ref{Sec:Results}, comparing them to the previously presented semi-analytical solutions. More precise ‘validity maps’ are then built and the findings discussed in the context of literature data. Then, in Section \ref{Sec:Hydrogen}, the study is particularised to the analysis of metals exposed to hydrogen-containing environments. As hydrogen reduces $J_{Ic}$, the influence of the environment in reducing sample size requirements and enabling micro-scale tests is assessed through dedicated `validity maps' that incorporate the $J_{Ic}$ sensitivity to the hydrogen content as an additional dimension. Finally, concluding remarks are provided in Section \ref{Sec:Conclusions}.


\section{Fundamental theory: HRR field and semi-analytical solutions}
\label{Sec:Analytical} 

In this Section, we provide a brief overview of the key theoretical concepts in our analysis. Specifically, in Section \ref{Sec:HRR}, we introduce the HRR singularity ahead of the crack tip, which serves as the fundamental basis for assessing the maximum valid $J$-dominance in the finite element (FE) simulations. Then, in Section \ref{Sec:Size}, we present a semi-analytical solution for the size requirement in fracture toughness tests, offering a rapid estimation of the maximum valid $J$-integral and minimum size requirements, which provide an initial basis for discussion and can be directly compared with FE simulations. 

\subsection{HRR singularity}
\label{Sec:HRR} 

The existence of the so-called HRR field, named after Hutchinson, Rice and Rosengren \cite{hutchinson_singular_1968,rice_plane_1968}, ensures $J$-dominance and similitude. Consider an elastic-plastic material following a power-law hardening relationship, such that the flow stress equals,
\begin{equation}\label{Eq:powerlaw}
    \sigma  = \sigma_Y \left(1 + \frac{\varepsilon^p}{\varepsilon_Y} \right)^N,
\end{equation}
where $\varepsilon^p$ is the effective plastic strain, $\varepsilon_Y=\sigma_Y/E$ is the initial yield strain, and $N$ is the strain hardening exponent, such that $N=0$ reduces Eq. (\ref{Eq:powerlaw}) to the case of an elastic-perfectly plastic solid.\\

It is convenient to introduce the crack tip opening displacement (CTOD), $\delta$, defined as the distance between the intersections of two 45° lines (forming a total angle of 90°) drawn backward from the crack tip with the deformed crack surface. The CTOD can be related to the $J$ integral by the expression \cite{shih_relationships_1981}
\begin{equation}\label{Eq:delta2}
\delta=d_n \frac{J}{\sigma_Y},
\end{equation}
where $d_n$ is a coefficient that strongly depends on the hardening exponent $N$ and is slightly influenced by other material properties such as $\sigma_Y$ and $E$. For an elastic-perfectly plastic solid, $d_n$ is approximately 1.0. For $N$ in the range of 0.05 to 0.33, $d_n$ varies between 0.2 and 0.7. Typically, for common structural steels, $d_n$ ranges from 0.4 to 0.8 \cite{shih_relationships_1981}.\\

Let us now introduce the scaling relationships resulting from the HRR singularity. For simplicity, we consider a uniaxial stress state. Following the power-law strain hardening in Eq. (\ref{Eq:powerlaw}), the strain energy density can be obtained as
\begin{equation}\label{Eq:energy1}
    \psi_s = \int_{0}^{\varepsilon} \sigma d \varepsilon= \frac{1}{N+1} \sigma_Y \varepsilon_Y \left(\frac{\sigma}{\sigma_Y}\right)^{1+\frac{1}{N}}.
\end{equation}

On the other hand, in a cracking solid, energy is released at the crack tip. This energy release can also be quantified in terms of its density, which can be approximated as
\begin{equation}\label{Eq:energy2}
    \psi_r = c \frac{J}{r},
\end{equation}
where $c$ is a constant and $r$ is the distance ahead of the crack tip. Combining Eqs. (\ref{Eq:energy1})-(\ref{Eq:energy2}) and noting that $\psi_s = \psi_r$ \cite{martinez2019mode}, we obtain
\begin{equation}\label{Eq:HRR_b}
    \left(\frac{\sigma}{\sigma_Y}\right)^{\frac{N+1}{N}} = \left(N+1\right) \frac{c}{\sigma_Y \varepsilon_Y} \frac{J}{r}.
\end{equation}

The above equation is generalised to multiaxial stress states following $J_2$ deformation theory and can be extended to the well-known HRR singularity
\begin{equation}\label{Eq:HRR}
    \sigma_{ij} \propto \left( \frac{J}{r}\right)^{\frac{N}{N+1}} \,\,\,\,\,\, \text{and} \,\,\,\,\,\, \varepsilon_{ij} \propto \left( \frac{J}{r}\right)^{\frac{1}{N+1}}.
\end{equation}

Eq. (\ref{Eq:HRR}) indicates that, in a log-log plot, the stress field exhibits a linear relationship with a slope of $N/(N+1)$, while the strain field follows a slope of $1/(N+1)$, as a function of the distance ahead of the crack tip. This relationship offers a straightforward criterion for evaluating the validity of $J$-dominance in fracture toughness testing. A numerical validation and discussion of the HRR singularity will be presented in Section \ref{Sec:Validation_HRR}.

\subsection{Specimen size requirements for a valid $J_{Ic}$ test}
\label{Sec:Size}

The existence of an HRR/$J$-field guarantees the \emph{priciple of similitude}; for any component, the stress state is equivalent for a given $J$, and uniquely characterised by this quantity. Thus, a critical $J$ value that is geometry-independent can be determined at the moment of failure, provided an HRR field exists \cite{begley_j-integral_1972,landes_effect_1972}. This has been widely exploited to characterise the fracture resistance of elastic-plastic materials that fail at conditions in which $K$-based characterisation is unfeasible. Consequently, efforts have been allocated to determine the regimes over which the HRR singularity is present and thus establish the associated  size requirements for $J_{Ic}$ testing (see the review by Zhu and Joyce \cite{zhu_review_2012} and Refs. therein). For the plain strain condition, a generalised form of the size requirement to ensure the existence of a HRR singularity and thus the validity of the $J_{Ic}$ test is given by
\begin{equation}\label{Eq:size3}
    a,(W-a) \geqslant M \left(\sigma_Y/E, N \right) \frac{J_{Ic}}{\sigma_Y},
\end{equation}
or equivalently according to Eq. (\ref{Eq:delta2})
\begin{equation}\label{Eq:size4}
    a,(W-a) \geqslant M \left(\sigma_Y/E, N \right) \frac{\delta}{d_n},
\end{equation}

Eq. (\ref{Eq:size3}) can be re-written as
\begin{equation}\label{Eq:size5}
    J_{Ic} \leqslant \frac{a,(W-a)}{M}{\sigma_Y},
\end{equation}
which enables the semi-analytically determination of the maximum valid $J$-integral, $J_\mathrm{max}$, in the fracture toughness test for a given yield strength and crack length or ligament size.\\

It remains to define the value of $M$ and various approaches have been taken in this regard. McMeeking \cite{mcmeeking_finite_1977} and Needleman and Tvergaard \cite{needleman_crack_1983} showed that the effect of finite strains is significant over a distance of approximately 2–3 times the crack tip opening displacement $\delta$. Therefore, one of the conditions for the validity of the HRR field requires
\begin{equation}\label{Eq:size_require1}
    R \geqslant 3 \delta,
\end{equation}
where $R$ is the radius of the HRR-dominated region. Bearing in mind that $d_n$ ranges from 0.4 to 0.8 for common structural steels, considering Eq. (\ref{Eq:delta2}) and selecting an intermediate value of $d_n=0.6$, one reaches
\begin{equation}\label{Eq:size_require2}
    R \geqslant 1.8 \frac{J_{Ic}}{\sigma_Y}.
\end{equation}

The second requirement is that $R$ must exceed the size of the fracture process zone, where the microscopic separation process occurs. The fracture process zone is typically smaller than the region of finite strain domination; e.g., in ductile metals fracture is due to nucleation, growth, and coalescence of micro-voids, with these processes occurring in the finite strain region. Hence, Eq. (\ref{Eq:size_require2}) is a suitable constraint also for the size of the fracture process zone size.\\ 

When the fully plastic condition is attained, the plastic zone extends across the entire uncracked ligament, and the length of the HRR domain $R$ becomes a fraction of $(W-a)$. Shih \cite{shih_requirements_1981,shih_j-dominance_1985} conducted extensive numerical studies on bend configurations and compact tension specimens and suggested that
\begin{equation}\label{Eq:R1}
    R \approx 0.07 (W-a).
\end{equation}

Combining Eqs. (\ref{Eq:size_require2}) and Eqs. (\ref{Eq:R1}), we have
\begin{equation}\label{Eq:M1}
    (W-a) \geqslant 25 \frac{J_{Ic}}{\sigma_Y},
\end{equation}
which corresponds to the value explained in Section \ref{Sec:Intro}. Note that for shallow-cracked specimens, the validity of the HRR field is primarily controlled by the crack length $a$ rather than the ligament \cite{chao_j-a2_1998,anderson_specimen_1991}. Consequently, the left-hand side of Eq. (\ref{Eq:M1}) should be replaced with the crack length $a$.\\ 

However, numerous studies have shown that $M=25$ is insufficient to fully satisfy the size requirements for $J_{Ic}$ testing, with the required value ranging from 40 to 250 depending on the specific material and specimen type \cite{yoshida_specimen_1989,kobayashi_fracture_2001,hutchinson_fundamentals_1983,anderson_specimen_1991}. For example, McMeeking and Parks \cite{mcmeeking_criteria_1979} and Shih \cite{shih_requirements_1981} compared the center-cracked panel (CCP) and the cracking bending bar (CBB) and concluded that CCP specimens require a significantly larger $M$ than CBB specimens. This is because, for the CCP specimen, the crack tip constraint is lower under tensile loading, leading to early plastic zone expansion and reduced HRR dominance. The uncracked ligament has significant plastic deformation, often forming 45° shear bands that alter the crack propagation path and facilitate large plastic straining. According to McMeeking and Parks \cite{mcmeeking_criteria_1979}, for the CCP specimen, the relationship between the radius $R$ and the ligament $(W-a)$ can be expressed as
\begin{equation}\label{Eq:R2}
    R \approx 0.01 (W-a),
\end{equation}
which leads to 
\begin{equation}\label{Eq:M2}
    (W-a) \geqslant 180 \frac{J_{Ic}}{\sigma_Y}.
\end{equation}

The value $M=25$ was incorporated into ASTM E813-81 \cite{noauthor_astm_1982} and remains widely used within the community. However, this version was withdrawn in 1997 and replaced by the standards ASTM E1737-96 \cite{noauthor_astm_1996} and ASTM E1820-96 \cite{noauthor_astm_1996-1}, which introduced a distinction in size requirements based on the $J$-measurement under stable crack growth ($J_{Q}$) and at unstable fracture ($J_{Qc}$). When applying $J_{Q}$ as $J_{Ic}$, the size requirements remain the same as those in Eq. (\ref{Eq:M1}) with $M=25$. However, when unstable fracture is involved, the recommended size requirements are given by
\begin{equation}\label{Eq:M3}
    a, (W-a) \geqslant 200 \frac{J_{Ic}}{\sigma_Y}.
\end{equation} 

In the most recent ASTM E1820-24 standard \cite{noauthor_astm_2024}, the size requirements were both relaxed, ultimately leading to the revised recommendations
\begin{equation}\label{Eq:M4}
    a, (W-a) \geqslant 10 \frac{J_{Ic}}{\sigma_Y},
\end{equation} 
for stable crack growth $J_{Q}=J_{Ic}$, and 
\begin{equation}\label{Eq:M5}
    a, (W-a) \geqslant 100 \frac{J_{Ic}}{\sigma_Y},
\end{equation} 
for unstable fracture $J_{Qc}=J_{Ic}$.\\

\begin{figure}[H]
\centering
\noindent\makebox[\textwidth]{%
\includegraphics[scale=0.45]{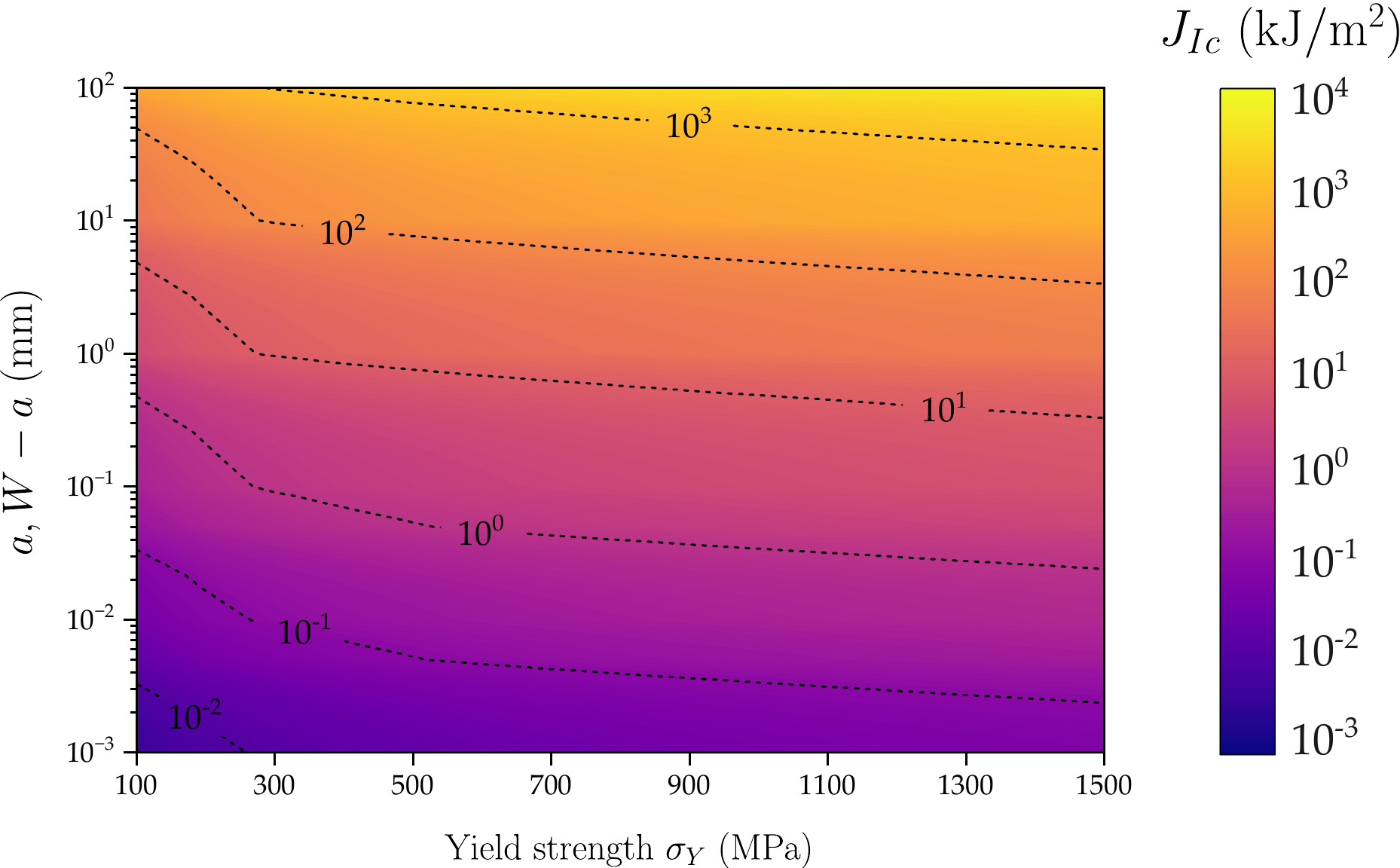}}
\caption{`Validity map' based on the simple semi-analytical equation (\ref{Eq:size5}), with $M=25$. The results show the minimum lengths of the crack and crack ligament as a function of the material yield strength $\sigma_Y$ and fracture toughness $J_{Ic}$.}
\label{fig:Map_analytical}
\end{figure}

Despite its importance in governing the allowable sample size, selecting the most relevant $M$ value is not straightforward, as it is material, geometry and fracture process dependent. To accurately establish the regimes of $J_{Ic}$ validity for the two paradigmatic micro-scale geometries shown in Fig. \ref{fig:Exp_photo}, finite element calculations are conducted for a wide range of geometries and material properties, and the outcome of these calculations will be compared to semi-analytical predictions with various $M$ values, so as to quantify the error associated with these more simplistic approaches. Nevertheless, first-order insight can be gained by building a `validity map' using Eq. (\ref{Eq:size5}). This is shown in Fig. \ref{fig:Map_analytical} where, for the widely used choice of $M=25$, the fracture toughness ($J_{Ic}$) is plotted as a function of the material yield strength and the minimum value of the relevant dimensions (crack and ligament lengths) that will deliver a valid test. In other words, the map uniquely relates a given material ($J_{Ic}$, $\sigma_Y$) to the minimum sample dimensions for the toughness estimate to be valid. As such, $J_{Ic}$ is equivalent to $J_\mathrm{max}$, the maximum valid $J$ value that can be obtained for a given sample ($a$, $W-a$) and material plastic behaviour ($\sigma_Y$). Notably, the influence of the strain hardening exponent is not considered in these simple semi-analytical estimates. 

The results shown in Fig. \ref{fig:Map_analytical} reveal the expected trends. For a fixed value of $J_{Ic}$, valid tests can be obtained with smaller samples if the yield strength is high, although the sensitivity is not that strong as $\sigma_Y$ is assumed to vary over a smaller range than $J_{Ic}$. The validity map shows that micro-meter size samples can only deliver quantitative fracture estimates for materials with fracture toughness below 0.1 kJ/m$^2$. This a range that spans the toughness of ceramics, rocks and cement, suggesting that micromechanical testing can only produce quantitative fracture estimates for very brittle materials. We shall refine these estimations below, with the aid of finite element analysis, and assess whether the embrittlement of metals due to their exposure to hydrogen can bring them into the realm of \emph{quantitative} micro-scale experiments.

\section{Modelling framework}
\label{Sec:Model} 

\begin{figure}[H]
\centering
\noindent\makebox[\textwidth]{%
\includegraphics[scale=0.55]{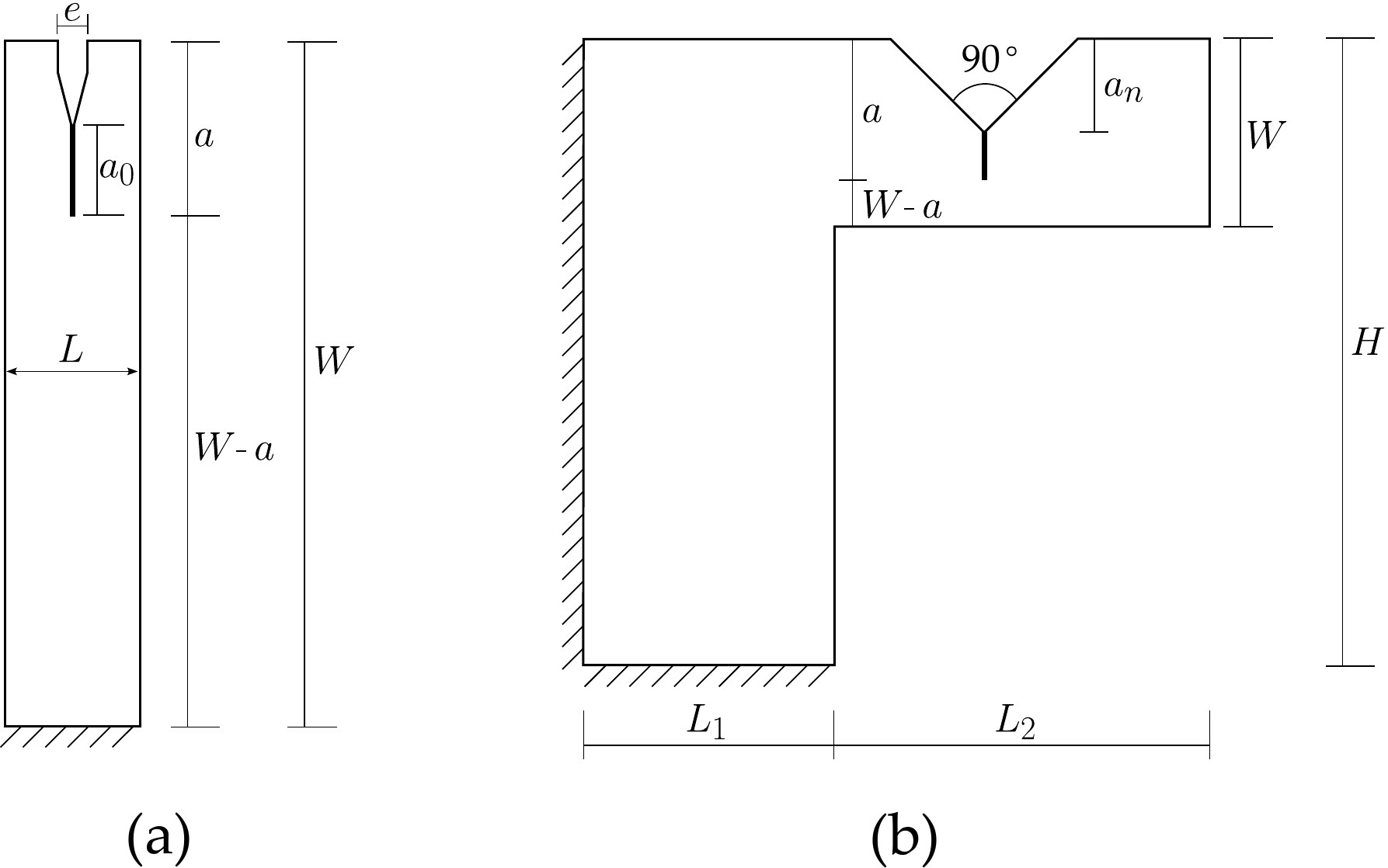}}
\caption{Geometric setup of the micro-scale samples considered in the finite element calculations: (a) double cantilever beam (DCB), and (b) notched cantilever beam specimen.}
\label{fig:DCB and bending}
\end{figure}

Finite element calculations are conducted to establish the relationship between the sample dimensions, the material plastic behaviour ($\sigma_Y$ and $N$) and the maximum valid critical $J$ value that can be measured for those conditions. For a given choice of material ($\sigma_Y$, $N$) and sample geometry, the distribution of tensile stresses ahead of the crack tip is plotted as a function of an increasing applied load ($J$ value), until an HRR field ceases to exist  - at this point: $J=J_\mathrm{max}$. The same analysis is repeated for dozens of geometries, with the validity maps being built using over 100 $J_\mathrm{max}$ data points.  The analyses are carried out for the two sample types commonly used in micro-scale testing: the double cantilever beam (DCB) specimen (Fig. \ref{fig:DCB and bending}a) and the notched cantilever beam specimen (Fig. \ref{fig:DCB and bending}b). The former is an example of a geometry that enables a long ligament length, and where the validity is likely limited by the crack size, with long cracks not being easy to introduce through FIB-milling due to material redeposition. On the other hand, the notched cantilever beam sample facilitates the milling of a long crack but is limited on the crack ligament length. Their geometries are shown in Fig. \ref{fig:DCB and bending}.

Typical micro-scale DCB experiments adopt a width-to-length ratio $L/W$ in the range 0.13-0.25, while the initial normalized crack length $a/W$ is relatively shallow, typically varying from 0.1 to 0.25 due to fabrication limitations \cite{sernicola_situ_2017,gavalda-diaz_basal_2021}. In this study, we set representative values of $L/W=0.25$, $a/W=0.125$, and $e/L=0.2$, implying that the validity of the test is mostly constrained by the crack length (as in lab experiments). For the notched cantilever beam specimen, we define $a_n/W=0.5$ and $a/W=0.6$, mimicking typical experimental configurations \cite{takahashi_direct_2016} and resulting in a dominant role of the uncracked ligament $(W-a)$ on the test validity (or lack thereof). Other dimensions, such as $L_1$, $L_2$, and $H$, do not play a primary role in the bending test, although the specimen height $H$ is typically much larger than $W$. Here, we set the representative values as $W/H=0.3$, $L_1/H=0.3$, and $L_2/H=0.7$, respectively. Plane strain conditions are assumed and only half of the DCB sample is modelled, taking advantage of symmetry.\\

\begin{figure}[H]
\centering
\noindent\makebox[\textwidth]{%
\includegraphics[scale=0.55]{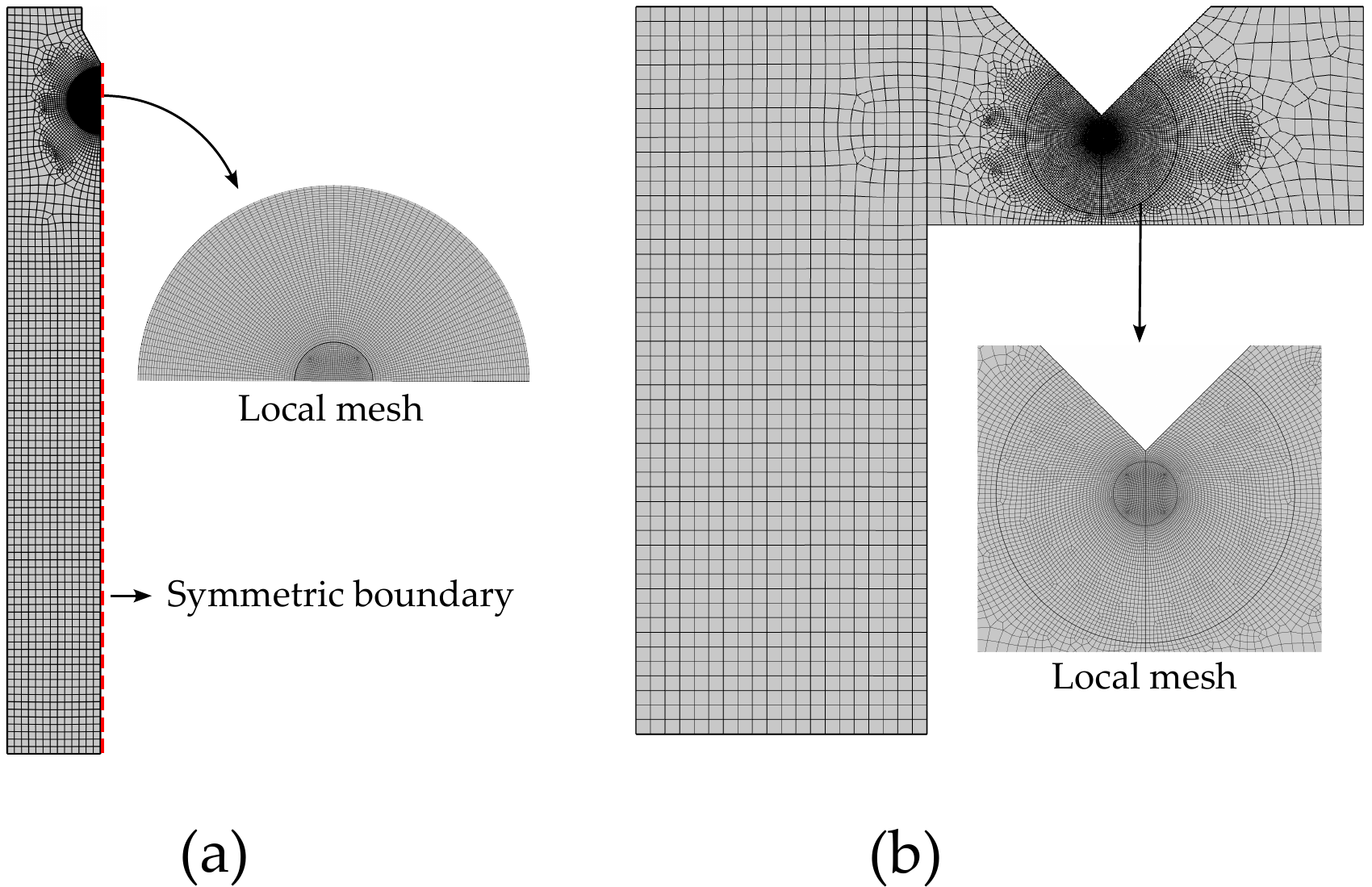}}
\caption{Finite element mesh of (a) a double cantilever beam (DCB) and (b) a notched cantilever beam specimen, with the mesh being refined near the crack tip.}
\label{fig:Mesh}
\end{figure}

Numerical simulations are conducted using the finite element package \texttt{COMSOL MULTIPHYSICS}. Quadratic quadrilateral serendipity elements with reduced integration are adopted for both geometries. Finite strains and rotations are considered. As shown in Fig. \ref{fig:Mesh}, the finite element mesh is refined near the crack/notch to appropriately resolve crack tip stresses. The resulting finite element mesh is the outcome of a mesh convergence study, which was conducted for representative geometries and material properties, including the limit cases. Consistent with the experimental conditions shown in Fig. \ref{fig:Exp_photo}, the DCB specimen is fixed along the bottom edge, while the notched cantilever beam specimen is constrained at both the left and bottom edges (Fig. \ref{fig:DCB and bending}). The numerical analyses are run in a load-controlled fashion, with a crack opening load being applied in the DCB and a downward load being applied on the edge of the cantilever beam, as in the experiments. Quasi-static conditions are assumed, with the load being increased progressively using very small load increments, to accurately capture the maximum load level at which the HRR field ceases to be valid. The relationship between the applied load and the value of the $J$-integral is obtained by using the contour integral method. The value of $J$ is obtained over multiple contours to ensure that the converged value is attained. \\

The material properties are varied along relevant ranges. The yield strength is varied from 100 MPa to 1500 MPa, encompassing the most commonly-used metallic materials, including high-strength steels, nickel-based alloys, and aluminum alloys. The crack length or ligament size ranges from 0.001 mm to 100 mm, covering a broad spectrum of small scale testing applications. Two different hardening exponents in Eq. (\ref{Eq:powerlaw}) are considered: $N=0.1$, representative of most low hardening materials such as carbon and low-alloy steels, and $N=0.3$, characteristic of high hardening materials like nickel-based alloys and copper-based alloys. It is expected that the higher hardening exponent $N=0.3$ will enhance the material’s resistance to localised plastic deformation near the crack tip, leading to a more constrained plastic zone and, consequently, a greater $J_\mathrm{max}$ (or a smaller minimum sample size, for a given $J_{Ic}$). However, this effect is not explicitly captured in the semi-analytical solution given by Eq. (\ref{Eq:size5}), where $M$ is a constant value independent from the strain hardening behaviour.

\section{Numerical results}
\label{Sec:Results} 

The computational framework is now used to assess the suitability of small scale testing for measuring fracture toughness. First, in Section \ref{Sec:Validation_HRR}, we showcase the ability of the model to resolve the HRR singularity field, using a representative geometry and material, and demonstrate how the maximum valid value of $J$-dominance, $J_\mathrm{max}$, is determined. Next, in Section \ref{Sec:Comparison}, we present representative comparisons between finite element simulations and semi-analytical solutions for different values of $M$. Finally, in Section \ref{Sec:Maps}, `validity maps' are generated through extensive FE analysis, showing how $J_\mathrm{max}$ varies as a function of both yield strength and crack length or ligament size. 


\subsection{Quantifying the HRR singularity regime}
\label{Sec:Validation_HRR} 

Let us begin by illustrating the ability of the model to resolve the expected crack tip behaviour and map regimes of $K$ and $J$ validity. To this end, the DCB sample is chosen as representative. The crack length is set to $a = 0.01$ mm, leading to a specimen geometry of $W = 0.08$ mm and $L = 0.02$ mm. The material parameters are defined as follows: Young's modulus $E = 210$ GPa, Poisson’s ratio $\nu = 0.3$, yield stress $\sigma_Y = 900$ MPa, and hardening exponent $N = 0.1$. Based on these parameters, the semi-analytical solution for the maximum value of $J$ at which the HRR field ceases to exist renders $J_\mathrm{max}=0.36$ $\mathrm{kJ/m^2}$, when $M = 25$.\\ 

\begin{figure}[H]
\centering
\noindent\makebox[\textwidth]{%
\includegraphics[scale=0.7]{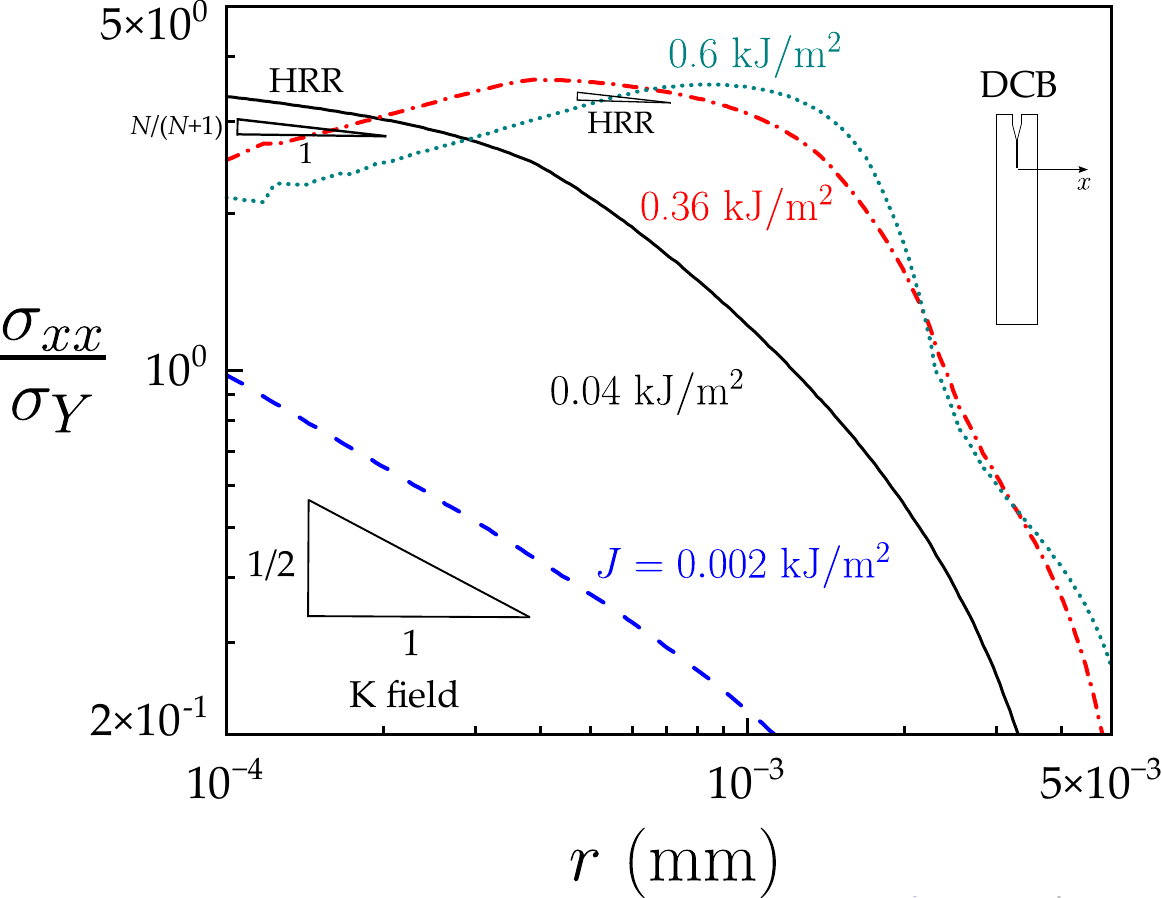}}
\caption{Crack tip stress distributions for various levels of the applied load. The results are computed for the DCB case and shown in a log-log plot, with the normalised tensile stress on the $y$-axis and the distance ahead of the crack $r$ on the $x$-axis. Results obtained for the extended crack plane (polar coordinate $\theta=0$) for a crack length $a = 0.01$ mm and material properties: $E = 210$ GPa, $\nu = 0.3$, $\sigma_Y = 900$ MPa, and $N = 0.1$.}
\label{fig:HRR1}
\end{figure}

The results obtained for different levels of the applied load are shown in Fig. \ref{fig:HRR1}, a log-log plot where the horizontal axis corresponds to the normalised distance ahead of the crack ($r/a$, based on a polar coordinate system centred at the crack tip), and the vertical axis corresponds to the normalised tensile (opening) stress, $\sigma_{xx}/\sigma_Y$. As shown for the $J=0.002 \, \mathrm{kJ/m^2}$ case (blue dashed curve), for very small applied loads the behaviour is almost purely elastic, with the stress field exhibiting a slope of 0.5, associated to the $K$-field singularity ($r^{-0.5}$). As the load is increased, the plastic zone increases in size and the distinct HRR singularity emerges, as characterised by a slope of $N/(N+1)$ (see Eq. (\ref{Eq:HRR})). This can be readily observed of the case of $J=0.04 \, \mathrm{kJ/m^2}$ (black solid curve), where the HRR regime is largest. As we keep ramping up the load, large strain plasticity and crack tip blunting start to dominate, resulting in a drop in the stress distribution and a smaller HRR regime. As discussed above, the semi-analytical expression predicts that the HRR domain will cease to exist when $J=0.36 \, \mathrm{kJ/m^2}$. The numerical results for that load level (dashed-dotted red line), still show a regime of HRR behaviour but this is relatively small. If the applied load is increased sufficiently, the HRR regime ceases to exist, as illustrated for a $J$ value of 0.6 $\mathrm{kJ/m^2}$ (green dotted curve). From a practical standpoint, we define the existence of the HRR regime (or lack thereof) from the stress distribution over the logarithmically transformed horizontal coordinate, $\log_{10}(r/a)$. Specifically, the HRR field is considered to have ceased to exist when the log-transformed length of the region over which the stress field follows the theoretical singularity falls below 0.05. By using this criterion, the maximum load level at which the HRR regime is still valid in this representative case is $J=0.52 \, \mathrm{kJ/m^2}$. This is a value slightly higher than the semi-analytical solution for $M=25$, with larger differences expected for materials with higher work hardening exponents.

\subsection{Representative comparisons with semi-analytical solutions}
\label{Sec:Comparison} 

The semi-analytical solution presented in Section \ref{Sec:Analytical} offers a simple, straightforward, and practical approach for estimating the size requirements of a valid fracture toughness for a given material. However, as previously discussed, while predictions are very sensitive to the value of $M$, there appears to be no clear consensus on its magnitude, and a constant value fails to reflect its high sensitivity to the strain hardening behaviour of the material. Thus, we proceed to compare the semi-analytical solutions with representative finite element simulations to quantitatively evaluate whether the chosen $M$ is sufficiently conservative across a wide range of yield strengths and varying hardening exponents.\\

In this sub-section, we examine two representative ligament (or crack) sizes in micro-scale testing: $a, (W-a) = 0.01$ mm and $a, (W-a) = 1$ mm. Note that both the ligament and crack sizes discussed here serve as the controlling factors in size requirements, depending on the specific specimen type (DCB vs cantilever beam). Consequently, larger ligament or crack sizes correspond to greater samples (overall). For each ligament (or crack) size, the yield strength is varied from 100 MPa to 1500 MPa, and the maximum valid $J$ value, $J_\mathrm{max}$, is computed following the procedure outlined in Section \ref{Sec:Validation_HRR}. \\

\begin{figure}[H]
\centering
\noindent\makebox[\textwidth]{%
\includegraphics[scale=0.4]{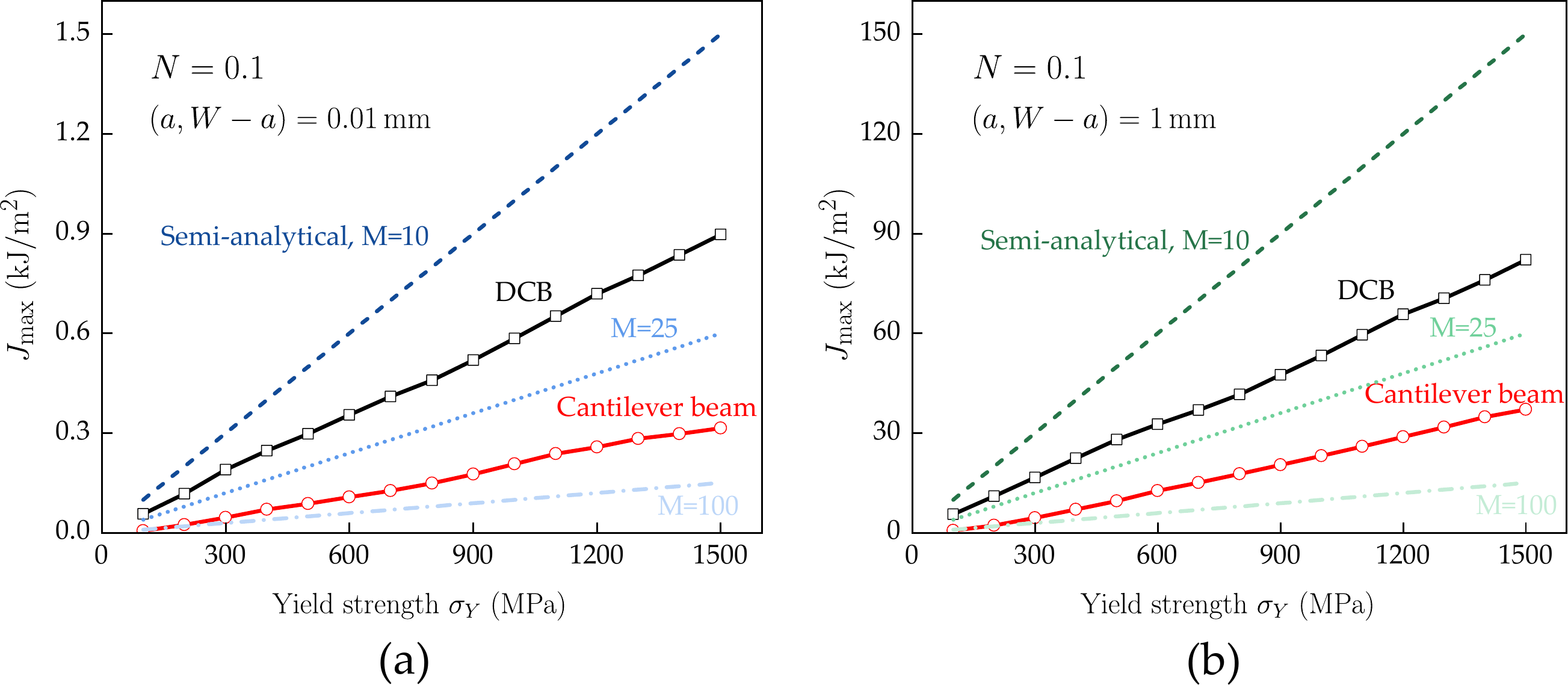}}
\caption{Finite element simulations of $J_\mathrm{max}$ as a function of yield strength for a hardening exponent of $N=0.1$ and ligament (crack) sizes of (a) 0.01 mm and (b) 1 mm. Results are compared with semi-analytical solutions for three representative values: $M=10$, $M=25$, and $M=100$.}
\label{fig:Comparison0.1}
\end{figure}

Consider first the results for a lower hardening exponent $N=0.1$. Fig. \ref{fig:Comparison0.1} shows the dependence of $J_\mathrm{max}$ on yield strength. For comparison, semi-analytical solutions using three different values of $M$ - namely $M=10$, $M=25$, and $M=100$ - are also presented. In general, numerical results show an approximately linear relationship between $J_\mathrm{max}$ and yield strength, consistent with the semi-analytical solution described in Eq. (\ref{Eq:size5}). Additionally, when the ligament (crack) size increases from 0.01 mm (Fig. \ref{fig:Comparison0.1}a) to 1 mm (Fig. \ref{fig:Comparison0.1}b), the predicted $J_\mathrm{max}$ increases by nearly two orders of magnitude, confirming a proportional relationship between $J_\mathrm{max}$ and ligament size. \\

\begin{figure}[H]
\centering
\noindent\makebox[\textwidth]{%
\includegraphics[scale=0.4]{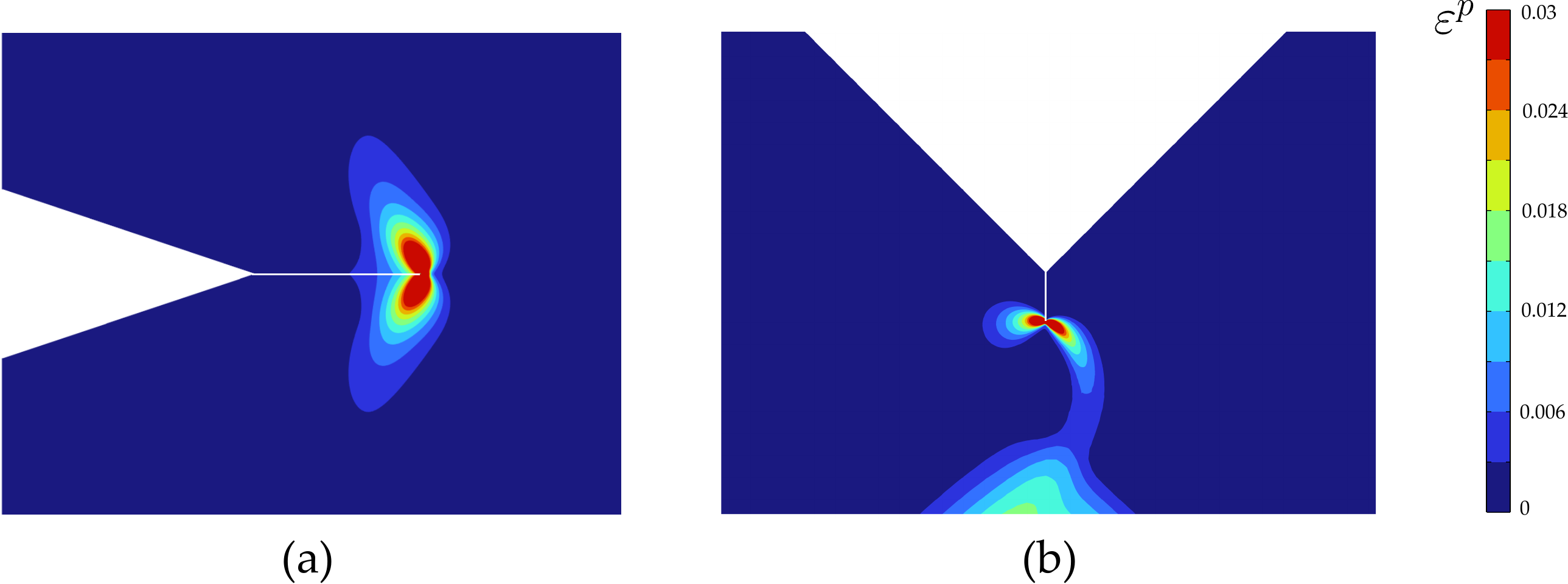}}
\caption{Contours of the effective plastic strain $\varepsilon^p$ for (a) DCB and (b) notched cantilever beam specimens with a hardening exponent of $N=0.1$, at $J=0.15$ $\mathrm{kJ/m^2}$, $\sigma_Y=900$ MPa, and $(a, W-a) = 0.01$ mm. }
\label{fig:EP}
\end{figure}

One interesting finding is that, for all cases considered, the simulated values of $J_\mathrm{max}$ for the DCB specimen (black curves) are higher than those for the notched cantilever beam specimen (red curves). This suggests that the size requirement for the notched cantilever beam specimen is more stringent than that for shallow-cracked specimens controlled by crack length (such as the DCB), a finding in agreement with previous observations \cite{chao_j-a2_1998}. This difference can be explained by looking at the equivalent plastic strain contours predicted, as given in Fig. \ref{fig:EP}. These results have been obtained for both DCB (Fig. \ref{fig:EP}a) and notched cantilever beam (Fig. \ref{fig:EP}b) specimens at $J = 0.15$ $\mathrm{kJ/m^2}$, $\sigma_Y = 900$ MPa, and $(a, W-a) = 0.01$ mm. At this $J$-integral value, the notched cantilever beam specimen is already very close to its $J_\mathrm{max}$ (0.18 $\mathrm{kJ/m^2}$), whereas the DCB specimen has a much higher $J_\mathrm{max}$ of 0.52 $\mathrm{kJ/m^2}$. From Fig. \ref{fig:EP}b, it can be observed that in the notched cantilever beam specimen, significant plastic strain develops at the bottom of the beam, propagating toward the crack tip plastic zone due to the limited ligament size. This rapid expansion of the plastic zone ultimately destabilises the $J$-dominance. In contrast, Fig. \ref{fig:EP}a shows that in the DCB specimen, the plasticity remains confined to a finite region, allowing $J$-dominance to be continuously maintained.

Based on the finite element simulation results, the semi-analytical solution with $M=10$ would underpredict the minimum size requirements for $J$-validity in both DCB and notched cantilever beam specimens. The commonly adopted value of $M=25$ provides a sufficiently conservative estimation for DCB tests but still overestimates $J_\mathrm{max}$ in the notched cantilever beam specimen. Finally, using $M=100$ ensures a fully conservative estimation for both specimen types, reinforcing its reliability as a safer criterion for determining size requirements, particularly for materials with a lower hardening exponent $N$.\\

\begin{figure}[H]
\centering
\noindent\makebox[\textwidth]{%
\includegraphics[scale=0.4]{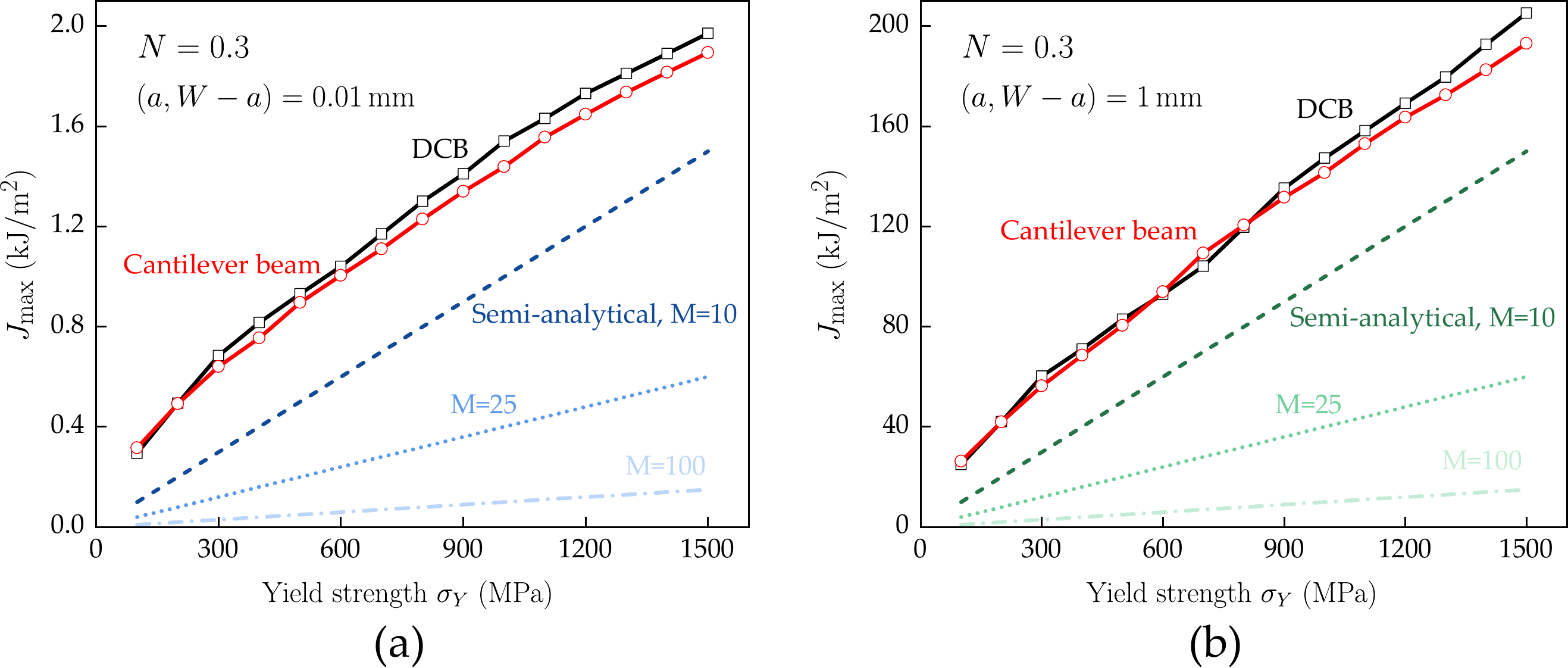}}
\caption{Finite element simulations of $J_\mathrm{max}$ as a function of yield strength for a hardening exponent of $N=0.3$ and ligament (crack) sizes of (a) 0.01 mm and (b) 1 mm. Results are compared with semi-analytical solutions for three representative values: $M=10$, $M=25$, and $M=100$.}
\label{fig:Comparison0.3}
\end{figure}

Finally, Fig. \ref{fig:Comparison0.3} shows the results obtained for a hardening exponent of $N=0.3$. Compared to Fig. \ref{fig:Comparison0.1}, it is evident that increasing the hardening exponent from $N=0.1$ to $N=0.3$ significantly raises the predicted $J_\mathrm{max}$ for both DCB and notched cantilever beam specimens. Moreover, the differences in $J_\mathrm{max}$ between the two specimen types become negligible at higher hardening exponents. With $N=0.3$, all investigated values of $M$ (10, 25 and 100) provide a conservative estimate for the size requirements and maximum valid $J$-integral $J_\mathrm{max}$. As in the $N=0.1$ case and as in Eq. (\ref{Eq:size5}), the numerical results in Fig. \ref{fig:Comparison0.3} exhibit an approximately linear relationship with yield strength.

\subsection{Validity maps relating yield strength, fracture toughness and sample dimensions}
\label{Sec:Maps} 

\begin{figure}[H]
\centering
\noindent\makebox[\textwidth]{%
\includegraphics[scale=0.35]{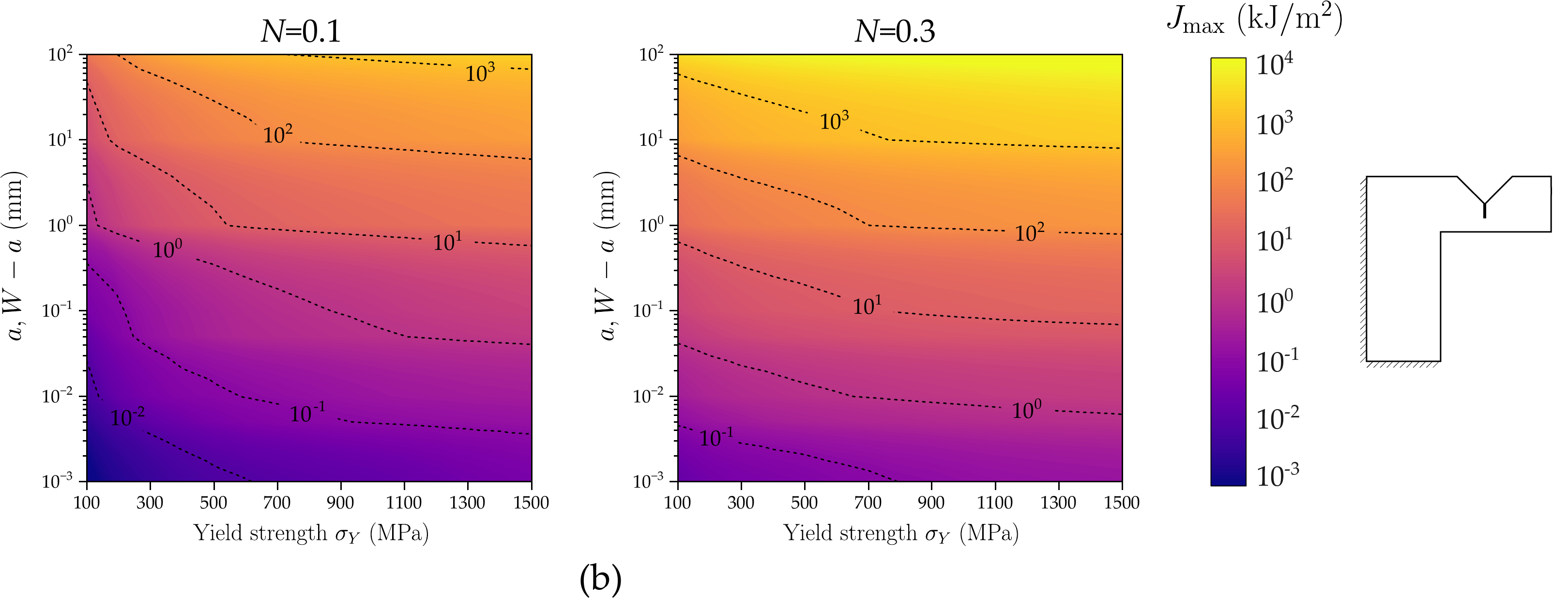}}
\caption{Maps showing finite element predictions of the maximum valid $J$-dominance $J_\mathrm{max}$ as a function of yield strength (x-axis) and ligament or crack size (y-axis) for (a) DCB specimen, and (b) notched cantilever beam specimen. Note that the value of $J_\mathrm{max}$ in these maps corresponds to the material toughness $J_{Ic}$ in Fig. \ref{fig:Map_analytical}, as determined using the semi-analytical solution.}
\label{fig:Map_NB}
\end{figure}

Following the procedure introduced in Section \ref{Sec:Validation_HRR} and Section \ref{Sec:Comparison}, extensive parametric studies are conducted to map the relationship between the maximum valid $J$ value ($J_\mathrm{max}$), the material yield strength and the relevant sample dimension (ligament or crack size). Results are shown in Fig. \ref{fig:Map_NB}. As expected, the colour gradient in the maps transitions from purple in the lower-left corner to yellow in the upper-right corner, indicating that as yield strength and specimen size increase, the maximum valid $J$-integral, $J_\mathrm{max}$, rises consistently. The maps also clearly show that a higher hardening exponent $N=0.3$ leads to greater $J_\mathrm{max}$ values compared to lower hardening levels, given the same yield strength and ligament or crack size. 

\begin{figure}[H]
\centering
\noindent\makebox[\textwidth]{%
\includegraphics[scale=0.35]{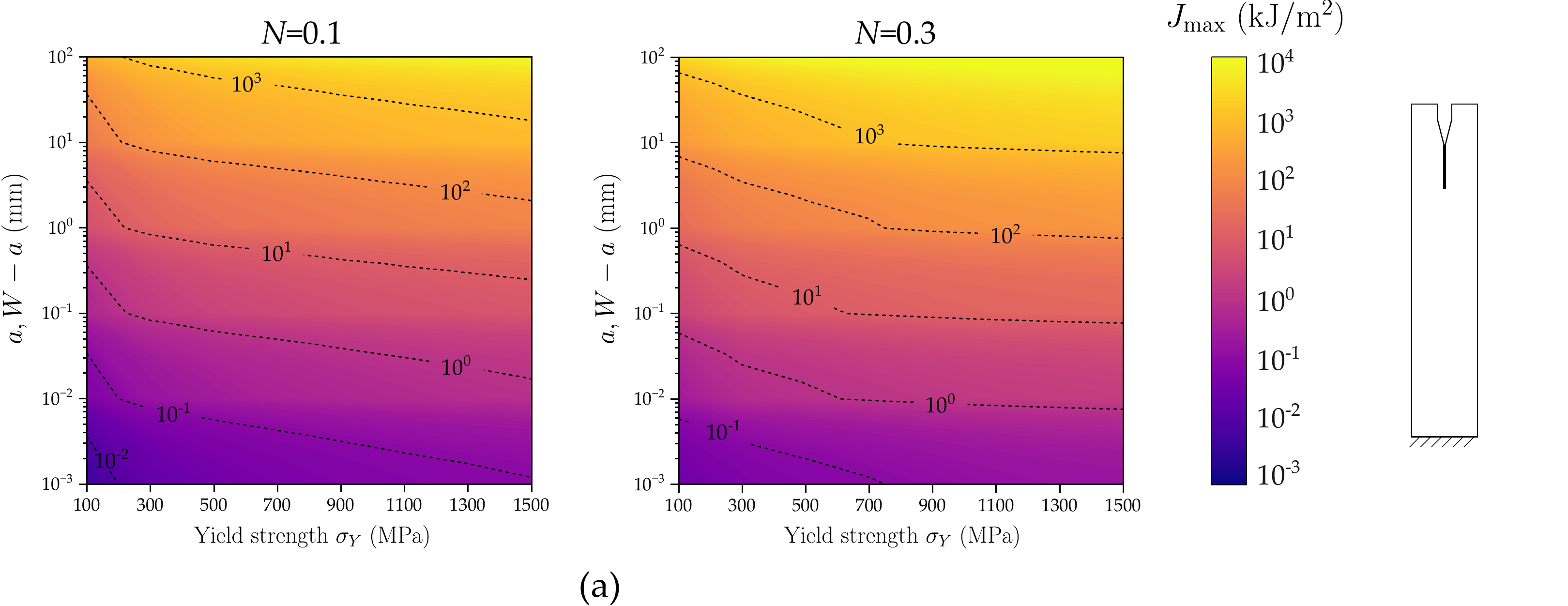}}
\label{fig:Map_DCB}
\end{figure}
\vspace{-1.2cm}

The trends are qualitatively similar to those obtained from the simple, semi-analytical equation (see Fig. \ref{fig:Map_analytical}). However, some quantitative differences emerge. For example, while Fig. \ref{fig:Map_analytical} suggested that the realm of micromechanical tests was limited to materials with toughnesses below 0.1 kJ/m$^2$, the numerical map reveals that both DCB and notched cantilever beam micromechanical tests can be used to quantitatively predict fracture in materials with toughness values on the order of $\sim 1$ kJ/m$^2$, for $N=0.3$ and relatively high strengths. This expands the applicability of micromechanical tests to the analysis of exceptionally brittle metals like beryllium and some polymer materials (PMMA, epoxy, polyester). However, most metallic materials have $J_{Ic}$ values at room temperature that significantly exceed that range. The differences between numerical and semi-analytical predictions will be examined in more detail and discussed below. 

\subsection{Discussion}
\label{Sec:discussions}

\begin{figure}[H]
\centering
\noindent\makebox[\textwidth]{%
\includegraphics[scale=0.35]{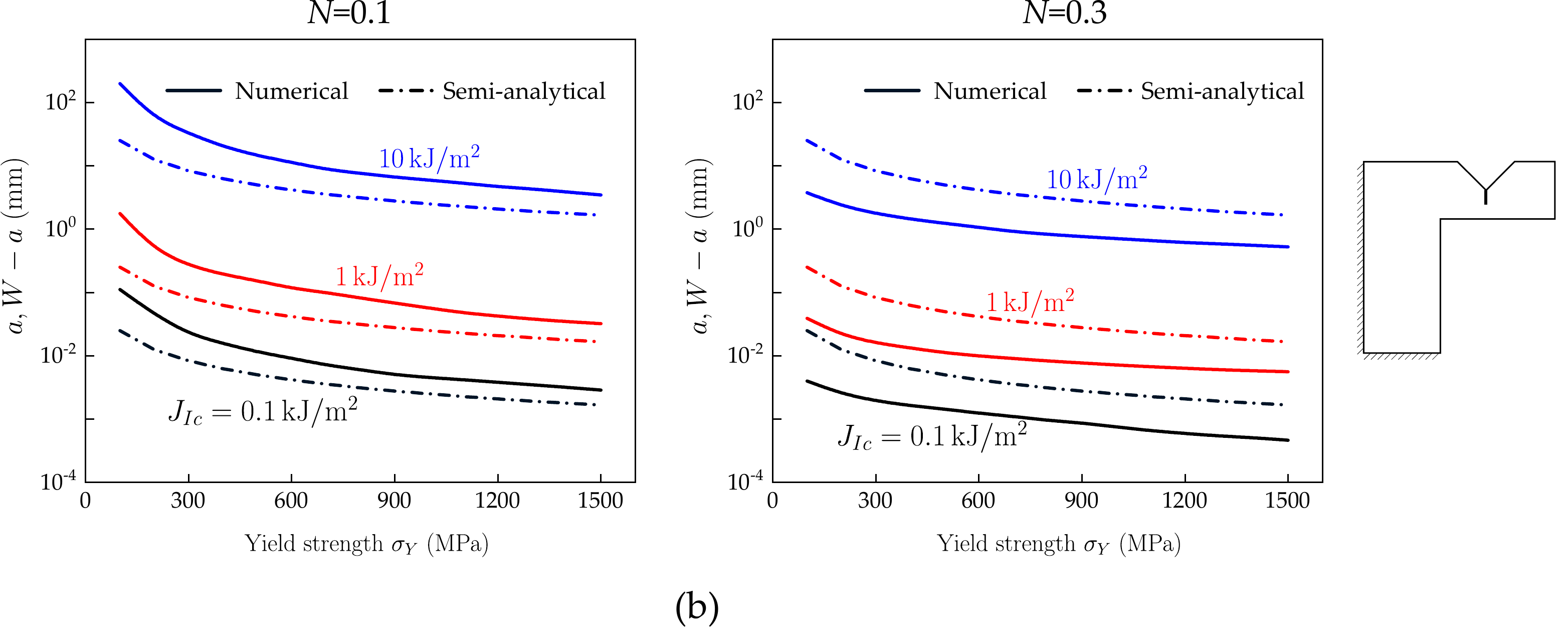}}
\caption{Comparison of the minimum size requirements for (a) DCB and (b) notched cantilever beam specimens, as predicted by numerical simulations and semi-analytical solutions with $M=25$, for three representative material toughness values: $J_{Ic}=0.1$ $\mathrm{kJ/m^2}$, $J_{Ic}=1$ $\mathrm{kJ/m^2}$, and $J_{Ic}=10$ $\mathrm{kJ/m^2}$.}
\label{fig:NBerror}
\end{figure}

We now compare the maps predicted by the numerical simulations (Section \ref{Sec:Results}) and the semi-analytical solutions (Section \ref{Sec:Analytical}). Using the maps shown in Fig. \ref{fig:Map_NB}, we can conveniently select multiple points with the same $J_\mathrm{max}$ value and investigate the minimum size requirement as a function of yield strength for different values of $J_\mathrm{max}$. Fig. \ref{fig:NBerror} shows the differences in size requirements between numerical and semi-analytical predictions for three representative material toughness values: $J_{Ic}=0.1$ $\mathrm{kJ/m^2}$, $J_{Ic}=1$ $\mathrm{kJ/m^2}$, and $J_{Ic}=10$ $\mathrm{kJ/m^2}$. Consistent with Fig. \ref{fig:Comparison0.1} and Fig. \ref{fig:Comparison0.3}, the semi-analytical solution with $M=25$ (dotted curves) predicts a more conservative (i.e., larger) size requirement for the DCB specimen, with this trend being particularly evident in materials with high strain hardening capacity. Additionally, the numerical predictions with $N=0.3$ show minimal dependence on specimen type, yielding nearly identical results for both DCB and notched cantilever beam specimens. However, when $N=0.1$, the semi-analytical solution predicts significantly lower size requirements for the notched cantilever beam specimen, especially for materials with lower yield strength (200 MPa or less), where the differences could be 5-8 times.\\ 

The differences between numerical simulations and semi-analytical solutions raise the question of whether the value $M$=10, as adopted in the latest ASTM standard \cite{noauthor_astm_2024}, or the widely used $M$=25 within the community, are sufficiently conservative. For high-hardening materials (i.e., $N\to 0.3$), both values appear to be conservative enough. However, the results shown in Figs. \ref{fig:Comparison0.1}, \ref{fig:Comparison0.3}, and \ref{fig:NBerror} suggest that while $M$=25 remains acceptable for DCB specimens with a hardening exponent of $N=0.1$, the use of bending (cantilever) specimen for small-scale testing of low-hardening materials requires more careful consideration of the ligament size. In summary, from an experimental perspective, $M$=10 is recommended for high-hardening materials, as per the ASTM standard. In contrast, for low-hardening materials -particularly in ligament-controlled cases - the use of $M=10$ should be revised. A more conservative value of $M=100$, which corresponds to the criterion for unstable fracture in the ASTM standard, is recommended.

\begin{figure}[H]
\centering
\noindent\makebox[\textwidth]{%
\includegraphics[scale=0.35]{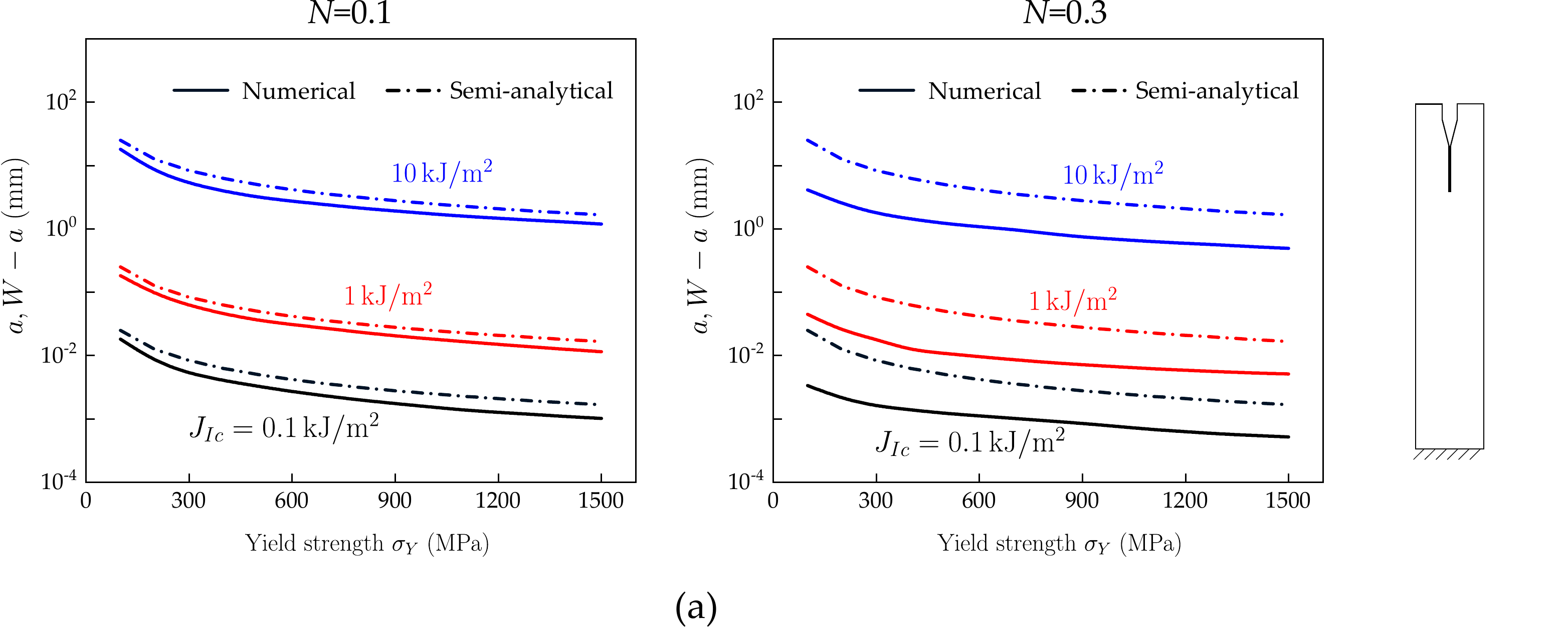}}
\label{fig:DCBerror}
\end{figure}
\vspace{-1.2cm}

Finally, we conclude this section by using our numerical maps to evaluate the validity of small scale fracture experiments from the literature. As shown in Fig. \ref{fig:NBerror}, for the DCB specimen, the semi-analytical results are more conservative for both low and high strain hardening materials, favouring this approach. One example of successful quantitative testing with micromechanical DCB samples are the experiments by Gavalda-Diaz \textit{et al.} on $\mathrm{Ti_3SiC_2}$  \cite{gavalda-diaz_basal_2021}, where the measured $J_{Ic}$ is approximately 0.044 $\mathrm{kJ/m^2}$, while the corresponding $J_\mathrm{max}$ obtained from Fig. \ref{fig:Map_NB} is 0.17 $\mathrm{kJ/m^2}$. Examples of valid notched cantilever beam tests include studies on NiTi \cite{haghgouyan2019fracture} and AISI 4340 steel \cite{brown1965study} using mm samples, and on silicon \cite{jaya2015can} and on specific planes in $\alpha$-iron \cite{snartland2020fracture}, using micrometer samples. Contrary to these successes, we also found studies conducted outside of the regimes of $J$-validity, such as recent microscale experiments on low-hardening tungsten \cite{bohnert2016fracture} and high-hardening copper \cite{hu2024microscopic}. In any case, it should be noted that the use of microscale experiments to predict the fracture behaviour of metals can bring other complications. For example, the assumption of a homogeneous material will break down when crystal anisotropy starts playing a role.\\ 

A further interesting observation is that for high strain hardening materials, the valid $J_{Ic}$ values obtained from cantilever specimens can sometimes exceed the semi-analytical solution while still remaining below the $J_\mathrm{max}$ derived from numerical maps. One example is the NiTi test in Ref. \cite{haghgouyan2019fracture}, where the measured $J_{Ic}$ is approximately 136 $\mathrm{kJ/m^2}$. For those conditions, the semi-analytical solution renders $J_\mathrm{max}=81.2$ $\mathrm{kJ/m^2}$, implying that the test is invalid. However, the numerical prediction for $J_\mathrm{max}$ in this test is 388.3 $\mathrm{kJ/m^2}$. This highlights the limitations of the semi-analytical solution in neglecting the influence of strain hardening behaviours, and emphasises the importance of using more accurate numerical maps. \\

\section{Validity maps for hydrogen-embrittled metals}
\label{Sec:Hydrogen}

Finally, we proceed to apply our numerical validity maps to metals exposed to hydrogen-containing environments, an area of growing interest within the small scale testing community \cite{takahashi_direct_2016,deng_-situ_2017,asadipoor_macro-_2020,tao_review_2023}. A primary objective of small-scale testing in hydrogen-containing environments is to evaluate the fracture strength within a localised region or even a specific microstructural feature or constituent susceptible to hydrogen accumulation (e.g., grain boundaries \cite{takahashi_direct_2016} and acicular ferrite \cite{costin_investigation_2016}). Although the mechanism of hydrogen embrittlement remains a topic of debate \cite{dear_mechanisms_2017,lynch_discussion_2019,guedes2020role}, it is widely observed that hydrogen ingress into metals significantly deteriorates fracture resistance \cite{Gangloff2003,IJHE2024,cupertino-malheiros_suitability_2024}. Thus, there is a need to gain microstrutural insight into the mechanisms and the notable reduction of $J_{Ic}$ observed in hydrogen environments should facilitate this. The following analysis aims to identify the regimes where quantitative microstructural insight into hydrogen embrittlement can be gained.\\ 

Fracture toughness and yield strength data for relevant materials in hydrogen gas (H$_2$) environments, denoted as $J_{Ic} (C)$ and $\sigma_Y(C)$, are collected and provided in Table \ref{tab:HE}, along with the corresponding values in air, denoted as $J_{Ic} (0)$ and $\sigma_Y(0)$. It can be observed that the fracture toughness decreases by a factor of 5 to 50. In contrast, the yield strength remains nearly unchanged in the presence of hydrogen, consistent with many other findings \cite{xu_tensile_2009,SanMarchi2012a}. Thus, a reduced size requirement can be expected as described by Eq. (\ref{Eq:size3}).

\begin{table} [H]
  \centering
  \begin{threeparttable}
  \caption{Fracture toughness and yield strength of various materials in air and hydrogen-containing environments (as characterised by the H$_2$ pressure $p_{\mathrm{H_2}}$). The hydrogen-relevant toughness $J_{Ic}(C)$ and the yield strength $\sigma_Y(C)$ are provided together with the their associated values in air ($J_{Ic}(0)$ and $\sigma_Y(0)$).}
    \label{tab:HE}
    \begin{tabular*}{\hsize}{@{}@{\extracolsep{\fill}}ccccccc@{}}
    \toprule
    Material & 
    \makecell{$\sigma_Y(0)$ \\ (MPa)} & 
    \makecell{$J_{Ic}(0)$ \\ (kJ/m$^2$)} & 
    \makecell{$p_{\mathrm{H_2}}$ \\ (MPa)} & 
    \makecell{$\sigma_Y(C)$ \\ (MPa)} & 
    \makecell{$J_{Ic}(C)$ \\ (kJ/m$^2$)} & 
    Reference \\
    \midrule
    Pure Ni & 200 & 232 & 83 & / & 49 & \cite{srinivas_fracture_1993,bechtle_grain-boundary_2009} \\
    X750 Ni alloy & 738 & 229 & 13.8 & 738 & 19 & \cite{symons_effect_1999} \\
    St35 Carbon Steel & 294 & 329 & 2 & 294 & 42 & \cite{steiner2023dvgw} \\
    A516-70 Carbon Steel & 372 & 121 & 34.5 & 365 & 36 & \cite{lam_literature_2009} \\
    L485 Steel & 527 & 1000 & 10 & 527 & 78 & \cite{steiner2023dvgw} \\
    X80 Steel & 580 & 438 & 103 & 593 & 9.5 & \cite{yang_fracture_2015,SanMarchi2012a} \\
    CrMo4130 Steel & 635 & 68 & 69 & 635 & 4.4 & \cite{SanMarchi2012a} \\ 
    H8 Steel & 790 & 450 & 19.5 & / & 40 & \cite{alvarez_hydrogen_2019} \\
    X100 Steel & 910 & 92 & 21 & 850 & 7 & \cite{ronevich_hydrogen-assisted_2021} \\
    HY130 & 940 & 150 & 69 & 940 & 2.5 & \cite{SanMarchi2012a} \\ 
    42CrMo4 Steel & 1023 & 89 & 20 & / & 50 & \cite{peral_effects_2019} \\
    CrMo4145 Steel & 1055 & 56 & 41 & 1055 & 1.6 & \cite{SanMarchi2012a} \\ 
    \bottomrule
    \end{tabular*}   
  \end{threeparttable}
\end{table}

\begin{figure}[H]
\centering
\noindent\makebox[\textwidth]{%
\includegraphics[scale=0.45]{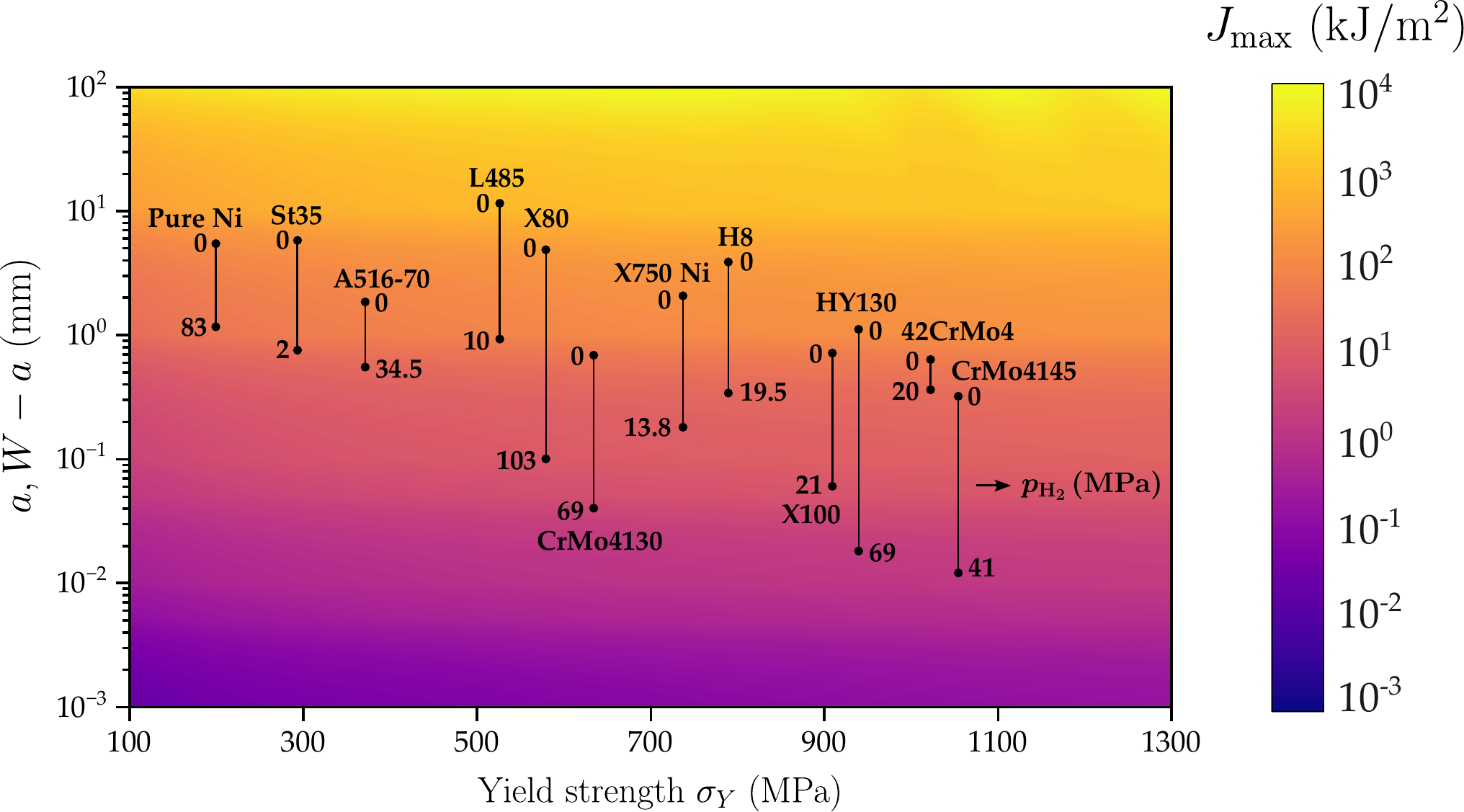}}
\caption{Validity map showing minimum ligament or crack size requirements (the solid black lines) for various materials in both air and hydrogen-containing environments based on finite element predictions with $N=0.3$. The numbers on the maps indicate hydrogen gas pressures collected from the literature and summarised in Table \ref{tab:HE}.}
\label{fig:Map_N0.3_H}
\end{figure}

We now integrate the fracture toughness data for both air and hydrogen gas environments, as summarised in Table \ref{tab:HE}, into the maps generated in Section \ref{Sec:Maps} to examine the changes in minimum size requirements due to hydrogen embrittlement. As an example, we use the finite element simulation results for the DCB specimen with a hardening exponent of $N=0.3$, as this choice enables us to identify the smallest size requirements and evaluate the constraints associated with size selection in small-scale testing for hydrogen-embrittled materials. The results, shown in Fig. \ref{fig:Map_N0.3_H}, include a series of black lines indicating the minimum size requirements for various materials under various hydrogen contents. As expected, the presence of hydrogen atoms reduces fracture toughness, thereby lowering the size requirements for the same material. Under high hydrogen gas pressure or concentration, the minimum size requirements can decrease by one to two orders of magnitude. For instance, CrMo4130 steel, which is widely used in aerospace and pipeline applications, experiences a reduction in the size requirement from $(a,W-a)\geqslant0.67$ mm to $(a,W-a)\geqslant0.04$ mm under a hydrogen gas pressure of 69 MPa. This finding suggests that small-scale testing techniques are more suitable for assessing hydrogen-assisted fracture behaviour than for fracture testing in air. However, as shown in Fig. \ref{fig:Map_N0.3_H}, for all investigated materials, the minimum required size $(a,W-a)$ remains above 0.01 mm. There could be significant challenges when aiming at fabricating samples with crack sizes and ligaments in the range of tens of microns, which would allow to investigate the hydrogen-embrittled local strength of specific microstructural features. 

Finally, it is important to note that Fig. \ref{fig:Map_N0.3_H} presents results for only a few representative materials. However, the methodology can be readily extended to any material of interest. In other words, the results and methodology provided enable a straightforward and practical approach for assessing the suitability of micro-scale testing in both air and agressive environments.

\section{Concluding remarks}
\label{Sec:Conclusions}

We have conducted both numerical and semi-analytical investigations to gain insight into the validity of  fracture toughness testing using small scale specimens. The simulations considered two typical micro-scale testing configurations: the double cantilever beam (DCB) and the notched cantilever beam specimen. A wide range of yield strengths, ranging from 100 MPa to 1500 MPa and encompassing most commonly used metallic materials, was examined, along with ligament sizes or crack lengths ranging from 0.001 to 100 mm. Two different hardening exponents, $N=0.1$ and $N=0.3$, were evaluated to investigate the influence of material hardening behaviour. Validity maps are built based on these investigations, offering a straightforward and practical approach for determining the minimum size requirements and assessing the applicability of quantitative fracture toughness measurements in small scale tests. The key findings include:

\begin{itemize} 
    \item The minimum size requirements $a,(W-a)$ and the maximum valid $J$-integral $J_{\mathrm{max}}$ are strongly dependent on the hardening exponent $N$ and are influenced by the specimen type. However, the effect of specimen type becomes negligible at higher hardening exponents $N$.
    \item For materials with a lower hardening exponent, neither the commonly used semi-analytical solution with $M=25$ nor the recommended criterion of $M=10$ in ASTM E1820 fully ensures compliance with minimum size requirements or the validity of fracture toughness tests. In such cases, a more conservative value of $M=100$ is recommended. In contrast, for high-hardening materials, $M=10$ remains sufficiently conservative.  
    \item The assessment of literature data reveals that the condition of $J$-validity is often violated, resulting in erroneous fracture estimates.
    \item The application of our analysis to hydrogen-embrittled metals suggests that it is possible to conduct quantitative micromechanical tests to gain insight into hydrogen embrittlement.
\end{itemize}


\section*{Declaration of Competing Interest}

The authors declare that they have no known competing financial interests or personal relationships that could have appeared to influence the work reported in this paper.

\section*{Acknowledgements}

C. Cui acknowledges financial support from the UKRI Horizon Europe Guarantee MSCA Postdoctoral Fellowship (grant EP/Y028236/1). E. Martínez-Pañeda was supported by a UKRI Future Leaders Fellowship (grant MR/V024124/1).

\section*{Data availability}

Data will be made available on request.




\end{document}